\documentclass[a4paper,12pt]{article}

\usepackage{pstricks}
\usepackage{color}
\usepackage{amssymb,amsmath,bm,bbm}
\usepackage{epsf,epsfig}
\usepackage{afterpage}
\usepackage{longtable}
\usepackage{cite}
\usepackage{latexsym,mathrsfs,dsfont}
\usepackage{graphicx}
\usepackage{url}
\usepackage{paralist}
\usepackage{bbold}
\usepackage{appendix}
\usepackage{hyperref}

\usepackage{soul}

\definecolor{dgreen}{rgb}{0.0, 0.5, 0.0}

\begin{document}

\hfill {\tt CERN-TH-2024-044}

\def\thefootnote{\fnsymbol{footnote}}

\begin{center}
\Large\bf\boldmath
\vspace*{1.cm} 
Exploring scalar contributions with \texorpdfstring{$K^+ \to \pi^+ \ell^+ \ell^-$}{K->pill}\unboldmath
\end{center}
\vspace{0.6cm}

\begin{center}
G.~D'Ambrosio$^{1}$\footnote{Electronic address: gdambros@na.infn.it}, 
A.M. ~Iyer$^{2}$\footnote{Electronic address: iyerabhishek@physics.iitd.ac.in}, F.~Mahmoudi$^{3,4,5}$\footnote{Electronic address: nazila@cern.ch}, 
S. Neshatpour$^{3}$\footnote{Electronic address: s.neshatpour@ip2i.in2p3.fr}\\
\vspace{0.6cm}
{\sl $^1$INFN-Sezione di Napoli, Complesso Universitario di Monte S. Angelo,\\ Via Cintia Edificio 6, 80126 Napoli, Italy}\\[0.4cm]
{\sl $^2$Department of Physics, Indian Institute of Technology Delhi,\\ Hauz Khas, New Delhi-110016, India}\\[0.4cm]
{\sl $^3$Universit\'e Claude Bernard Lyon 1, CNRS/IN2P3, \\
Institut de Physique des 2 Infinis de Lyon, UMR 5822, F-69622, Villeurbanne, France}\\[0.4cm]
{\sl $^4$Theoretical Physics Department, CERN, CH-1211 Geneva 23, Switzerland}\\[0.4cm]
{\sl $^5$Institut Universitaire de France (IUF), 75005 Paris, France }\\[0.4cm]
\end{center}
\renewcommand{\thefootnote}{\arabic{footnote}}
\setcounter{footnote}{0}

\vspace{1.cm}
\begin{abstract}
\noindent

The rare kaon decay $K^+ \to \pi^+\ell^+\ell^-$  offers insights into Standard Model (SM) physics and beyond. Driven by vector form factor in the SM, it can also probe non-standard contributions. In this letter we study the scalar contribution, $f_S$. Using differential decay width and Forward-Backward Asymmetry we propose a simultaneous fit to vector and scalar contributions which is necessary for a consistent analysis.  Novel bounds on $|f_S|$ are presented for the first time through a reinterpretation of the E865, NA48/2, and NA62 experimental results. The analysis results in the most precise bound $f_S < 7.9\times 10^{-6}$ at 90\% confidence level.

\end{abstract}

\thispagestyle{empty}

\clearpage
\section{Introduction}
Kaon physics has been fundamental for our understanding of the structure of weak interactions: discovery of the GIM mechanism, $CP$ violation, $P$ violation, etc.
There are several ongoing efforts for a deeper understanding of rare kaon decays: They include processes like $K_L\rightarrow\pi^0\nu\bar{\nu}$, $K^+\rightarrow\pi^+\nu\bar{\nu}$~\cite{KOTO:2020prk,NA62:2021zjw}, and $K^+ \to \pi^+ \ell^+ \ell^-$ among others.
The interest in semi-leptonic $K^+ \to \pi^+ \ell^+ \ell^-$ decay is to extract short-distance information:  lepton flavour universality violation test~\cite{Crivellin:2016vjc,DAmbrosio:2022kvb,HIKE:2022qra,Ahdida:2023okr,HIKE:2023ext,DAmbrosio:2023irq} and other short-distance probes like $P$ and $CP$ violation~\cite{DAmbrosio:1996lam,DAmbrosio:1998gur}. Chiral Perturbation Theory (ChPT)~\cite{Gilman:1979ud,Ecker:1987qi,Ecker:1987hd,DAmbrosio:1994fgc,DAmbrosio:1998gur,Cirigliano:2011ny,Ananthanarayan:2012hu,DAmbrosio:2018ytt} is the appropriate theory framework for describing this decay. Within the Standard Model (SM), this decay at short distances is induced at the loop level and predominantly occurs via single virtual photon exchange. Due to its suppression in the SM it offers a glimpse into different types of short-distance physics.

In this letter, we study the scalar contribution to $K^+ \to \pi^+ \ell^+ \ell^-$. 
It also affects processes such as $K_{L,S}\rightarrow\mu^+\mu^-$, $K_L\rightarrow \pi^0 \ell^+\ell^-$ in addition to $K^+ \to \pi^+ \ell^+ \ell^- $. 
The existing datasets from the electron and the muon channels~\cite{E865:1999ker,NA482:2009pfe,NA482:2010zrc} and the ongoing measurements of $K^+\to\pi^+\mu^+\mu^-$ decay at NA62~\cite{NA62:2022qes} make it an exciting probe to test the limits of the scalar operators.  
We address the fact that the only existing bound on scalar contributions from the E865 experiment from 1999 is derived from the branching ratio.
While the Forward-Backward Asymmetry $(A_{\rm FB})$ in these decays vanishes in the SM \cite{Ecker:1991ru,Ecker:1987hd,Gao:2003wy,Chen:2003nz}, the presence of scalar interactions introduces non-zero $A_{\rm FB}$, making it a powerful probe of New Physics.
Typical scenarios that can give rise to such effects have been studied for example in~\cite{Dedes:2003kp, Demir:2003bv,Isidori:2006qy}.
Although $A_{\rm FB}$ serves as a robust probe, its effectiveness can be influenced by the lepton mass, leading to suppression in the electron mode, where direct measurements are currently lacking.
Alternatively, the branching ratio (BR) offers another avenue for investigating scalar interactions. While slightly less sensitive than $A_{\rm FB}$ in the muon channel, the branching ratio serves as a primary probe in the electron channel, providing insights into the presence of scalar contributions~\cite{E865:1999ker}.

We propose a more thorough investigation into the scalar contribution by 
conducting a simultaneous fit to both the vector and the scalar contributions using the differential decay width bins, both when the $A_{\rm FB}$ measurement is included or not. Specifically, when conducting a combined investigation with $A_{\rm FB}$, which only exhibits non-zero values in the presence of scalar contribution, the inclusion of the scalar form factor in the fit is mandatory in order to have a consistent study.
We obtain bounds on scalar contributions from the different available experimental datasets.

After discussing the framework in section~\ref{sec:framework}, in section~\ref{sec:BR_AFB}
through the examination of the $A_{\rm FB}$ and branching ratio, we find an estimate of the sensitivity to scalar contributions. In section~\ref{sec:dGammadz_AFB} we obtain a bound from our proposed fit to data where in addition to the vector contributions, scalar contributions are taken into account and we summarise our results in section~\ref{sec:conclusions}.

\section{Framework}\label{sec:framework}
The amplitude of the $K^+ \to \pi^+ \ell^+ \ell^-$ decay,  when  taking into account only the vector and scalar interactions, as denoted by the form factors  $f_V$, and $f_S$ respectively, can be written as~\cite{Beder:1975ph,E865:1999ker}
\begin{align}\label{eq:amplitude}
{\cal M}=\frac{\alpha G_F}{4\pi}f_V(z) P^\mu \bar{\ell}\gamma_\mu \ell + G_F M_K f_S \bar\ell \ell \,,
\end{align}
where $P = p_K+p_\pi$
and $q=p_K-p_\pi$, 
with $p_K$ and $p_\pi$ the momenta of the kaon and the pion, respectively and the dilepton invariant mass squared can be written as $q^2 = z M_K^2$.

Given the above amplitude, the double-differential decay width, in terms of the vector and scalar form factors, is expressed as~\cite{Chen:2003nz,Gao:2003wy}
\begin{align}\label{eq:2diffWidth}
\frac{d^2\Gamma}{dz\,d\!\cos\theta}
&= \frac{G_F^2 M_K^5}{2^8\pi^3} \beta_\ell\, \lambda^{1/2}(z) 
 \times \left\{\left|f_V\right|^2\frac{\alpha^2}{16\pi^2}
\lambda (z)(1-\beta_\ell^2\cos^2\theta)+\left|f_S\right|^2z\beta_\ell^2 \right. \\[-6pt]\nonumber
&\qquad\qquad\qquad\qquad\qquad \left.+{\rm Re}(f_V^{\ast}f_S)\frac{\alpha \, r_\ell}{\pi} \beta_\ell \lambda^{1/2} (z)\cos\theta
\right\} \,, 
\end{align}
where $\theta$ is the angle between the negatively charged lepton and the kaon in the dilepton rest frame, $r_{\ell} = m_\ell/M_K$, $r_{\pi} = m_\pi/M_K$, $\beta_\ell = \sqrt{1-4r_\ell^2/z}$, and $\lambda(z)\equiv \lambda(1,z,r_\pi^2)$ is the K\"all\`en function.  

The familiar $z$-spectrum is recovered by integrating over $\cos\theta$
\begin{align}\label{eq:dGdz}
\frac{d\Gamma}{dz} =  \frac{2}{3}\frac{G_F^2 M_K^5}{2^8\pi^3} \beta_\ell \lambda^{1/2}(z) \times\left\{ \left|f_V\right|^2 2 \frac{\alpha^2}{16\pi^2}\lambda(z)\Big(1+2\frac{r_\ell^2}{z}\Big) \left|f_S\right|^2 3\, z\beta_\ell^2 
\right\}\,,
\end{align}
which upon further integration with respect to $z$ yields the branching ratio.

Another interesting observable is obtained by considering the angular behavior of the decay, with the Forward-Backward Asymmetry defined as
\begin{align}
A_{\rm FB}(z)=\frac{\int^1_0 \left(\frac{d\Gamma}{dzd\!\cos\theta}\right)d\!\cos\theta-\int^0_{-1}
\left(\frac{d\Gamma}{dzd\!\cos\theta}\right)d\!\cos\theta}{\int^1_0 \left(\frac{d\Gamma}
{dzd\!\cos\theta}\right)d\!\cos\theta+\int^0_{-1}
\left(\frac{d\Gamma}{dzd\!\cos\theta}\right)d\!\cos\theta}\,.
\end{align}
Considering Eq.~(\ref{eq:2diffWidth}), we have \cite{Gao:2003wy,Chen:2003nz}
\begin{align}\label{eq:AFB}A_{\rm FB}(z) = \left. \frac{\alpha G_F^2M_K^5}{2^8\pi^4}r_\ell\,\beta_{\ell}^2(z)\lambda(z)
{\rm Re}\left(f_V^* f_S\right) \!
\middle/ \!\left(\frac{d\Gamma(z)}{dz}\right) \right.\!\!,
\end{align}
which is non-zero only in case vector and scalar contributions are simultaneously present.

In the SM, the $K^+ \to \pi^+ \ell^+ \ell^-$ decay is completely governed by the vector form factor $f_V(z)$ which can be described as a linear contribution in $z$ accompanied by the unitarity loop correction~\cite{DAmbrosio:1998gur}\footnote{\label{Note1}The vector form factor has been described with several formulations, from a simplistic model with only a linear parameterisation, $f_V = f_0(1 + \delta z)$, to more complicated models~\cite{DAmbrosio:1998gur,Friot:2004yr,ColuccioLeskow:2016noe, Dubnickova:2006mk}. In this study, for $f_V$ we consider the ``Linear + Chiral'' description from Ref.~\cite{DAmbrosio:1998gur} as given in Eq.~(\ref{eq:fV_Lin_Chiral}), where $f_V$ here in terms of $W(z)$ of Ref.~\cite{DAmbrosio:1998gur} is given by $f_V = W(z)/(G_F M_K^2)$.}, expressed as 
\begin{align}\label{eq:fV_Lin_Chiral}
f_V(z) = a_+ + b_+z + V^{\pi\pi}(z).
\end{align}
Here, $V^{\pi\pi}(z)$ accounts for the pion loop contribution calculated at ${\cal O}(p^6)$ in ChPT~\cite{DAmbrosio:1998gur}. The parameters $a_+$ and $b_+$ are considered as phenomenological constants, typically extracted from experimental data. Nevertheless, recent advancements have been achieved in theoretical calculations concerning these parameters (e.g., see~\cite{DAmbrosio:2018ytt,DAmbrosio:2019xph,RBC:2022ddw}). 
On the other hand, the scalar form factor is highly suppressed and negligible in the SM. 
Regarding the available data, precise measurements for the differential decay width distributions of $K^\pm\to \pi^\pm \ell^+ \ell^-$ decay have been conducted since the initial observation of $K^+\to\pi^+e^+e^-$~\cite{Bloch:1974ua} at CERN. In the electron channel,  the most events have been observed by BNL-E865~\cite{E865:1999ker} and NA48/2~\cite{NA482:2009pfe}, while for the muon channel, there are results from NA48/2~\cite{NA482:2010zrc} and more recently NA62~\cite{NA62:2022qes}.

To investigate scalar contributions we consider both the branching ratio and the Forward-Backward Asymmetry.
The latter is clearly dependent on scalar contributions, where a non-vanishing $A_{\rm FB}$ necessitates non-zero scalar contributions. However, due to its proportionality to the lepton mass ($r_\ell = m_\ell/M_K$), it is highly suppressed in the electron channel. This suppression does not apply to the branching ratio which is obtained by integrating the differential decay width, Eq.~(\ref{eq:dGdz}), over $z$. 
Section~\ref{sec:BR_AFB} provides further elaboration on the approach.

A more concrete analysis assuming the presence of scalar contributions in addition to the vector form factors is given in  section~\ref{sec:dGammadz_AFB}. This choice is strongly motivated by the possible non-zero value of $A_{\rm FB}$.
The measurement of the $A_{\rm FB}$ makes it mandatory to have a three parameter fit in general.
While an experimental measurement of the $A_{\rm FB}$ is extremely difficult owing to the electron mass suppression, we can extract a more consistent bound on the scalar contribution by means of a three-parameter fit.

\begin{table*}[t]
\centering
\scalebox{0.9}{
\begin{tabular}{|c|c|c|c|c|c|c|}
\cline{1-3} \cline{5-7}
\multicolumn{3}{|c|}{\rule{0pt}{3ex}$(K^+\to\pi^+\mu^+\mu^-)$} & $\phantom{\qquad}$ & \multicolumn{3}{|c|}{$(K^+\to\pi^+e^+e^-)$}\\
\cline{1-3} \cline{5-7}
\multicolumn{7}{c}{}\\[-12pt]
\cline{1-3} \cline{5-7}
NA48/2         & exp & $|f_S|<$ & \qquad & E865~~\;         & exp & $|f_S|<$ \\
\cline{1-3} \cline{5-7}
BR           & \rule{0pt}{3ex}$(9.62 \pm 0.21) \times 10^{-8} $ & $ 1.0 \times 10^{-4} $ & & BR           & $(2.988 \pm 0.040) \times 10^{-7} $ & $ 6.8 \times 10^{-5} $\\
$A_{\rm FB}$ & $ (-2.4 \pm 1.8)  \times 10^{-2} $ & $ 4.2 \times 10^{-5} $ & & $A_{\rm FB}$ & -- & -- \\
\cline{1-3} \cline{5-7}
\multicolumn{7}{c}{}\\[-12pt]
\cline{1-3} \cline{5-7}
NA62~~\;         & exp & $|f_S|<$  & & NA48/2         & exp & $|f_S|<$\\
\cline{1-3} \cline{5-7}
BR           & \rule{0pt}{3ex}$(9.16 \pm 0.06) \times 10^{-8} $ & $ 5.6 \times 10^{-5} $ & & BR           & $\;(3.14 \pm 0.04) \times 10^{-7}\;$ & $ 6.8 \times 10^{-5} $\\
$A_{\rm FB}$ & $ (0.0 \pm 0.7)  \times 10^{-2} $ & $ 7.7 \times 10^{-6} $ & & $A_{\rm FB}$ & -- & --\\
\cline{1-3} \cline{5-7}
\end{tabular}
}
\caption{Bound on $|f_S|$ at 90\% CL, from $A_{\rm FB}$ and the uncertainty of the branching ratio. In each panel, the last column corresponds to the upper bound obtained from the experimental measurement of the column to its left. For the electron channel, there are no measurements available for the Forward-Backward Asymmetry.}
\label{tab:fS_from_AFB_BR}
\end{table*}

\section{Branching ratio vs. Forward-Backward Asymmetry}\label{sec:BR_AFB}
Historically, the scalar contributions were constrained by the branching ratio~\cite{E865:1999ker} and $A_{\rm FB}$~\cite{NA482:2010zrc}. 
The muon mode enjoys a model-independent measurement of the branching ratio. In this case, we have two independent sources for estimating
the sensitivity on scalar contributions: 1)~The measured branching ratio,  2)~The measurement of $A_{\rm FB}$.
On the other hand, there is no model-independent measurement of the branching ratio for the electron mode. The existing method for measuring the branching ratio for the electron mode is by assuming a model-specific vector form factor~(see footnote~\ref{Note1}).
There is only the branching fraction which allows for a measure of the sensitivity of the scalar contributions and there are no measurements on $A_{\rm FB}$.

For the branching ratio, to constrain $f_S$, we evaluate the permissible contributions from scalar interactions at 90\% confidence level (CL), considering the uncertainty in the measured branching ratio of $K^+\to\pi^+\ell^+\ell^-$. Regarding $A_{\rm FB}$, we examine the experimental results and derive the corresponding 90\% CL upper bound on $f_S$ employing the relation in Eq.~(\ref{eq:AFB}). Experimental data for  $A_{\rm FB}$ is, however, as mentioned before, only available for the muon channel. 
The bounds on $f_S$ that we derived from both the BR and the $A_{\rm FB}$ measurements are presented in Table~\ref{tab:fS_from_AFB_BR}, considering various experimental measurements. In each panel of Table~\ref{tab:fS_from_AFB_BR}, the second column corresponds to the experimental value, while the last column denotes the upper bound on $|f_S|$ that we obtain. 

Notice that the bound from the NA62 measurement of $A_{\rm FB}$ is approximately seven times stronger than the one  from the branching ratio as measured by the same experiment. However, it is important to note that this constraint may not apply to scenarios involving lepton flavour universality violating scalar contributions. Regarding the electron mode, current constraints are solely derived from the BR measurements, utilising either the NA48/2~\cite{NA482:2009pfe} or E865~\cite{E865:1999ker} results. 
Our result is in agreement with the upper bound given by the E865 experiment~\cite{E865:1999ker}.
While the NA62 collaboration has yet to measure the electron mode, assuming a similar enhancement as observed in the muon channel compared to NA48/2 (a factor of 3), the upper limit for $|f_S|$ in the electron channel could potentially decrease to $\sim 4 \times 10^{-5}$.

\section{Bound on \texorpdfstring{$\boldsymbol{f_S}$}{fS} from three-parameter fit}\label{sec:dGammadz_AFB}
The experimentally determined values of $a_+$ and $b_+$ are obtained by analysing $d\Gamma/dz$ data, assuming only vector contributions. It is possible to extend this analysis to include scalar contributions and re-evaluate the fit for $a_+$, $b_+$, and $f_S$. For the muon channel, measurements of $A_{\rm FB}$ can also be incorporated into the analysis.

In the previous section, it is practically assumed that the size of $|f_S|$ is small compared to the vector form factor. In this section with a three-parameter fit, we abandon this assumption to obtain a more solid evaluation of the scalar contributions.

In Table~\ref{tab:abfS_fit} we give the 90\% CL upper bound on $|f_S|$, obtained from the three-parameter fit to $a_+,b_+$, and $f_S$ with the different datasets~\cite{Mahmoudi:2007vz,Mahmoudi:2008tp,Mahmoudi:2009zz,Neshatpour:2022fak}. The $d\Gamma/dz$ data of the NA48/2 and NA62 measurements are available on the HEPData repository~\cite{hepdata.35365,hepdata.69636,hepdata.135498}.
The second (third) column corresponds to the fit considering $d\Gamma/dz$ bins, while excluding (including) $A_{\rm FB}$.
When $A_{\rm FB}$ is not included, the results are similar to what one gets from the branching ratio in the previous section. This is expected as BR effectively encapsulates the information from the differential decay width bins. Nonetheless, it would be interesting to see experiments also explore a three-parameter model, including $f_S$ in addition to $a_+$ and $b_+$ when analysing the data. On the other hand, the bound from the fit including both the $d\Gamma/dz$ and $A_{\rm FB}$ (last column in Table~\ref{tab:abfS_fit}) is very similar to the bound obtained from only $A_{\rm FB}$ (as given in Table~\ref{tab:fS_from_AFB_BR}). This is due to the fact that the bound from $A_{\rm FB}$ is far stronger than the decay width and adding the latter does not offer much further information. The consistency between the fit and the constraints from the preceding section on $f_S$ is reassuring and justifies the assumption of the dominance of the vector form factor made in the previous section.

\begin{table}
\centering
\begin{tabular}{|l||c||c|}\hline
& \multicolumn{1}{|c||}{$d\Gamma/dz$} & \multicolumn{1}{|c|}{$d\Gamma\!/dz + A_{\rm FB}$ }\\\hline \hline 
$K\to \pi ee$  & $|f_S| <$ & $|f_S| <$ \\\hline
E865\phantom{~\cite{E865:1999ker}}   &  $8.0 \times 10^{-5}$ & -- \\
NA48/2\phantom{~\cite{NA482:2009pfe}}  &  $4.0 \times 10^{-5}$ & -- \\\hline \hline
$K\to \pi \mu\mu$  & $|f_S| <$ & $|f_S| <$  \\\hline
NA48/2\phantom{~\cite{NA482:2010zrc}}  & $10.0 \times 10^{-5}$ & $4.1 \times 10^{-5}$
\\
NA62\phantom{~\cite{NA62:2022qes}}   & $9.0 \times 10^{-5}$  & $7.9 \times 10^{-6}$ \\\hline
\end{tabular}
\caption{Upper bound for $|f_S|$ at 90\% CL, from the three parameter fit to $a,b$ and $f_S$ using various datasets.  For the relevant inputs regarding the theoretical calculations we have considered PDG 2022~\cite{ParticleDataGroup:2022pth}, and the external parameters $\alpha_+,\beta_+$ are taken from~\cite{DAmbrosio:2022jmd}, in agreement with NA62~\cite{NA62:2022qes}.}
\label{tab:abfS_fit}
\end{table}

\begin{figure*}[t]
\centering
\includegraphics[width=0.49\linewidth]{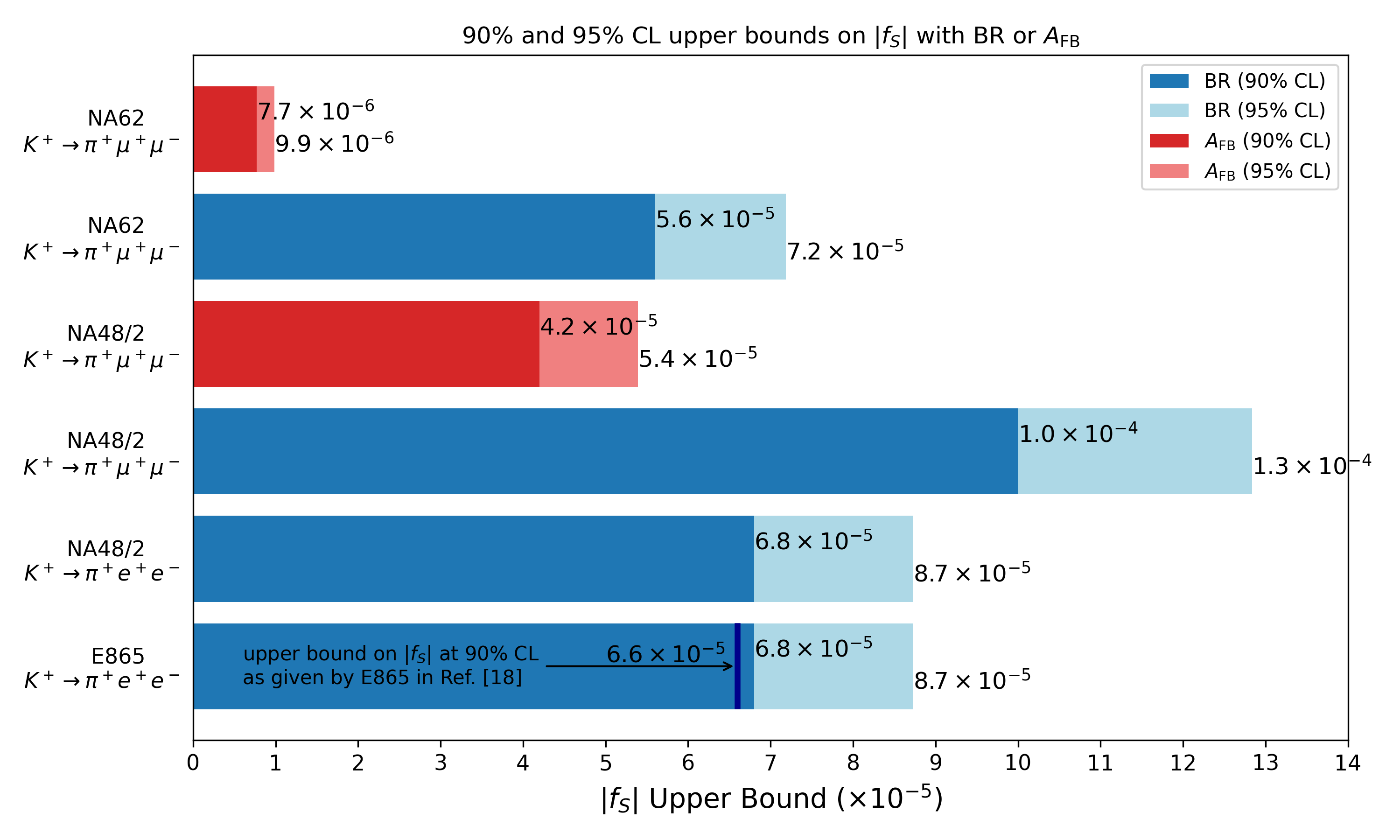} 
\includegraphics[width=0.49\linewidth]{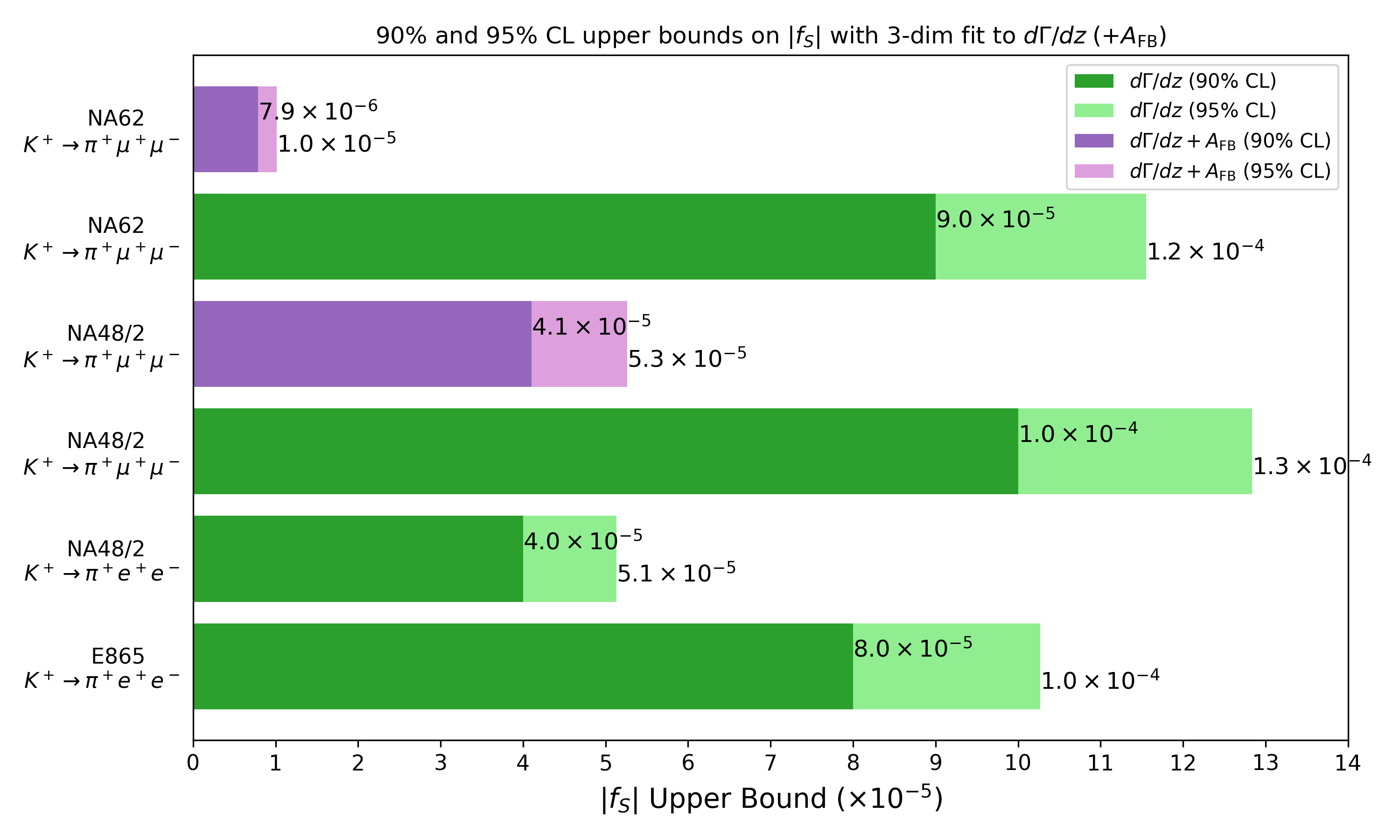}
\caption{The 90\% and 95\% CL upper bound on $|f_S|$ obtained by our analysis of different experimental datasets.
Left: bound from BR or $A_{\rm FB}$ as given in Table~\ref{tab:fS_from_AFB_BR}. Right: bound from 3-dim. fit to $f_V$ and $f_S$ as given in Table~\ref{tab:abfS_fit}. The vertical black line in the left plot corresponds to the only existing experimental upper bound from E865~\cite{E865:1999ker}. 
}
\label{fig:BR_AFB_comparison_3dfit}
\end{figure*}

\section{Summary}\label{sec:conclusions}
In this letter, we analysed the data on the $K^+\to\pi^+\ell^+\ell^-$ decay and constrained scalar contributions, $f_S$. 
Using BR or $A_{\rm FB}$ we obtain 90\% upper bounds on $f_S$ as given in Table~\ref{tab:fS_from_AFB_BR}. 
The shortcoming of this simple approach is the assumption of the relative smallness of $f_S$ compared to $f_V$. For a concrete and precise bound on $f_S$, we proposed a new approach to analyse the experimental data via a 3-dimensional simultaneous fit to $f_V$ and $f_S$ (Table~\ref{tab:abfS_fit}). This method is necessary to have a consistent analysis of the data, especially given the direct measurement of $A_{\rm FB}$. It is particularly crucial if a non-zero $A_{\rm FB}$ is measured, as is the case for the NA48/2 measurement of the muon mode.

A comparison between the two methods of sections~\ref{sec:BR_AFB} and~\ref{sec:dGammadz_AFB} is depicted in the left and right plots of Fig.~\ref{fig:BR_AFB_comparison_3dfit}, respectively. The only upper bound available in the literature, which is obtained using the  method of section~\ref{sec:BR_AFB} via BR, is indicated by the vertical black line in the left plot as given by~\cite{E865:1999ker}.
All the bounds represented by the colored bars are derived in this paper through the analysis of various experimental datasets. The slight difference of our upper bound obtained via the BR result of the E865~\cite{E865:1999ker}experiment compared to the value given in Ref.~\cite{E865:1999ker} is expected due to updated input parameters.

The most precise limit on $f_S$ that we obtain is $7.9\times 10^{-6}$ at 90\% CL which is about one order of magnitude stronger than the bound given in Ref.~\cite{E865:1999ker} by the E865 experiment. Our upper bound arises from the 3-dimensional fit to NA62 data when $A_{\rm FB}$ is included, highlighting the potential of the latter to probe scalar interactions. It would be interesting to analyse $A_{\rm FB}$ in smaller bins, as it could enhance our ability to further scrutinise scalar contributions.  
As the NA62 collaboration is planning to have an analysis of the full dataset of the $K^+ \to \pi^+ \ell^+\ell^-$ decay, this consideration will be particularly pertinent to further constrain the size of the scalar contributions.

\section*{Acknowledgments}
We appreciate the valuable input from E.~Goudzovski, M.~Koval, C.~Lazzeroni, and G.~Ruggiero of the NA62 collaboration. The stimulating discussions and correspondence have enhanced our comprehension concerning the data and their treatment.
This research was funded in part by the National Research Agency (ANR) under project ANR-21-CE31-0002-01. AMI would like to acknowledge the generous support by SERB India through the project SRG/2022/001003. AMI would also like to thank the French Institute- Embassy of France in India for facilitating research trip to IP2I Lyon in December 2022.
AMI would also like to thank the hospitality of INFN Sezione di Napoli and IP2I Lyon during the academic visit in December 2023.
GD was supported in part by the INFN research initiative Exploring New Physics~(ENP).

\bibliographystyle{JHEP} 
\bibliography{kaon_scalar}

\end{document}